\documentclass[journal,10pt]{IEEEtran}
\normalsize

\usepackage{cite}
\usepackage{graphicx}
\usepackage[cmex10]{amsmath}
\usepackage{amssymb}
\usepackage{booktabs}
\usepackage{threeparttable}
\usepackage{algorithm}
\usepackage{algpseudocode}
\usepackage{amsfonts}
\usepackage{xcolor}
\usepackage{color}

\usepackage{tabularx} %

\usepackage{caption}

\usepackage{wrapfig}

\usepackage{enumitem}
\usepackage{framed}

\usepackage{dblfloatfix} %

\usepackage{soul}

\usepackage{cancel} %

\usepackage{verbatim}%

\usepackage{multirow}

\usepackage{makecell}

\usepackage{empheq} %

\usepackage{underoverlap} %

\usepackage{hyperref}
\hypersetup{pdftex,colorlinks=true,allcolors=black}

\usepackage{calc}

\usepackage{textcomp}
\def\BibTeX{{\rm B\kern-.05em{\sc i\kern-.025em b}\kern-.08em
		T\kern-.1667em\lower.7ex\hbox{E}\kern-.125emX}}

\newcommand{\textbb}[1]{{\color{blue}#1}}

\newcommand{\textrr}[1]{{\color{red} #1 }}

\renewcommand{\deg}{^{\circ}}

\newcommand{\myDFTexp}[2]{ e^{ -\mathsf{j} \frac{2\pi(#1\cdot #2)}{N} } }

\begin{document}

\title{{Green Joint Communications and Sensing Employing Analog Multi-Beam Antenna Arrays}}

\author{
	Kai Wu,
	 J. Andrew Zhang,~\IEEEmembership{Senior Member,~IEEE}, %
	Xiaojing Huang,~\IEEEmembership{Senior Member,~IEEE},\\
	Robert W. Heath Jr.,~\IEEEmembership{Fellow,~IEEE}, and Y. Jay Guo,~\IEEEmembership{Fellow,~IEEE}
	
\vspace{-0.5cm}
}

\maketitle

\begin{abstract}	
	Joint communications and sensing (JCAS) is potentially a hallmark technology for the sixth generation mobile network (6G). Most existing JCAS designs are based on digital arrays, analog arrays with tunable phase shifters, or hybrid arrays, which are effective but are generally complicated to design and power inefficient.
	This article introduces the energy-efficient and easy-to-design multi-beam antenna arrays (MBAAs) for JCAS.
	 Using pre-designed and fixed analog devices, such as lens or Butler matrix, MBAA can simultaneously steer multiple beams yet with negligible power consumption compared with other techniques. 
Moreover, MBAAs enable flexible beam synthesis, accurate angle-of-arrival estimation, and easy handling/utilization of the beam squint effect. All these features have not been well captured by the JACS community yet. To promote the awareness of them, we intuitively illustrate them and also  
exploit them for constructing a multi-beam JCAS framework. 
Finally, the challenges and opportunities are discussed to foster the development of green JCAS systems.

\end{abstract}

\begin{IEEEkeywords}%
	Integrated sensing and communications (ISAC), joint communications and sensing (JCAS), multi-beam antenna arrays, beam steering, beam selection, angle-of-arrival estimation
\end{IEEEkeywords}

\section{Introduction}\label{sec: introduction}
Joint communications and sensing (JCAS), also known as integrated sensing and communications (ISAC), has recently attracted enormous attention from academia and industry \cite{ISAC_cui2021integrating,FanLiu_overview2020TCOM,JCAS_paul2016survey}. 
Performing communications and sensing with a single set of spectral, waveform and transceiver
resources, JCAS is promising to substantially improve the spectral and energy efficiency of future intelligent wireless systems whose intelligence will inevitably rely on data acquisition via sensing and communications (data transferring).

One of the most studied topics is JCAS beamforming for steering dual-functional beams, some for communications and others for sensing. 
Massive multiple-input multiple-output (MIMO) arrays
are extensively considered for JCAS beamforming, 
given the crucial role of massive MIMO in future millimeter-wave (mmWave) and terahertz (THz) communications and sensing  \cite{THz_sensingImagingLocal_comMag2020,HybridMIMO_switchesPSs2016}.
Moreover, as it can be seen from the recent JCAS overview/survey articles \cite{FanLiu_overview2020TCOM,JCAS_paul2016survey}, the majority of JCAS beamforming designs are based on the massive MIMO arrays with tunable phase shifters (PSs). 
{The PSs have limited phase resolution and constant amplitudes, which pose challenges to finding the optimal phase configuration of a massive MIMO array for JCAS beamforming.}

The same problem has also been experienced with massive MIMO communications in recent years. A viable solution developed by the communication community is using multi-beam antenna arrays (MBAAs) for analog beamforming \cite{Jay_Butler6G_overview,Lens_energyEfficient_ComMag2018,KaiWu_WIPT_LAA,Kai_WidebandEffect2019JSTSP,zeng2016millimeter}.
Such antenna arrays can simultaneously steer multiple fixed beams 
using pre-designed and fixed analog devices, such as lens or Butler matrix \cite{Jay_Butler6G_overview}. {The beamforming features of MBAAs will be introduced shortly in Section \ref{sec: multi-beam antenna arrays}.}
Their power consumption is negligible, as compared to the phased array with tunable PSs \cite{Lens_energyEfficient_ComMag2018}. 
Moreover, the multi-beam nature of the array can greatly simplify signal processing tasks, such as beam synthesis and angle estimation etc. \cite{KaiWu_WIPT_LAA,Kai_WidebandEffect2019JSTSP}. However, these benefits of MBAAs have not been captured and appreciated by the JCAS community yet.

This article aims to promote the awareness of MBAAs in the JCAS community, potentially inspiring greener JCAS solutions. 
We start by depicting the multi-beam function of MBAAs in Section \ref{sec: multi-beam antenna arrays}. We then review in Section \ref{sec: relevan designs} several signal processing designs/features, which, although developed for communications originally, are highly useful to JCAS as well. 
To demonstrate this, we jointly exploit those handy designs/features for advancing a recently developed multi-beam JCAS framework \cite{Andrew_Multibeam2019TVT}. This will be elaborated in Section \ref{sec: multi-beam JCAS framework}. 
Applying a relatively new antenna array to JCAS inevitably incurs challenges. We highlight some critical ones in Section \ref{sec: challenges of multi-beam antenna arrays}, along with potential solutions and research opportunities.

{The main contributions of the article include:

\begin{enumerate}
	\item We illustrate the mechanism and key features of the MBAA beamforming and compare MBAAs with the phase shifter-based massive MIMO arrays in terms of power consumption. 
	
	\item We intuitively introduce several interesting designs employing MBAAs in JCAS-critical topics, including beam synthesis, direction estimation and beam squint effect. 
		 
	\item We demonstrate the flexibility of applying MBAAs to a JCAS transmitter by extending a recently develop multi-beam JCAS framework \cite{Andrew_Multibeam2019TVT}. In particular, we illustrate both regular and random transmitting schemes and highlight their pros and cons, in terms of receiving complexity, power-angle relationship and communication secrecy. 
	
	\item We also discuss open issues of applying MBAAs to JCAS and link them with prior designs which, though developed for other applications, may enlighten potential solutions. 
\end{enumerate}}

\begin{figure*}[!t]
	\centering
	{\includegraphics{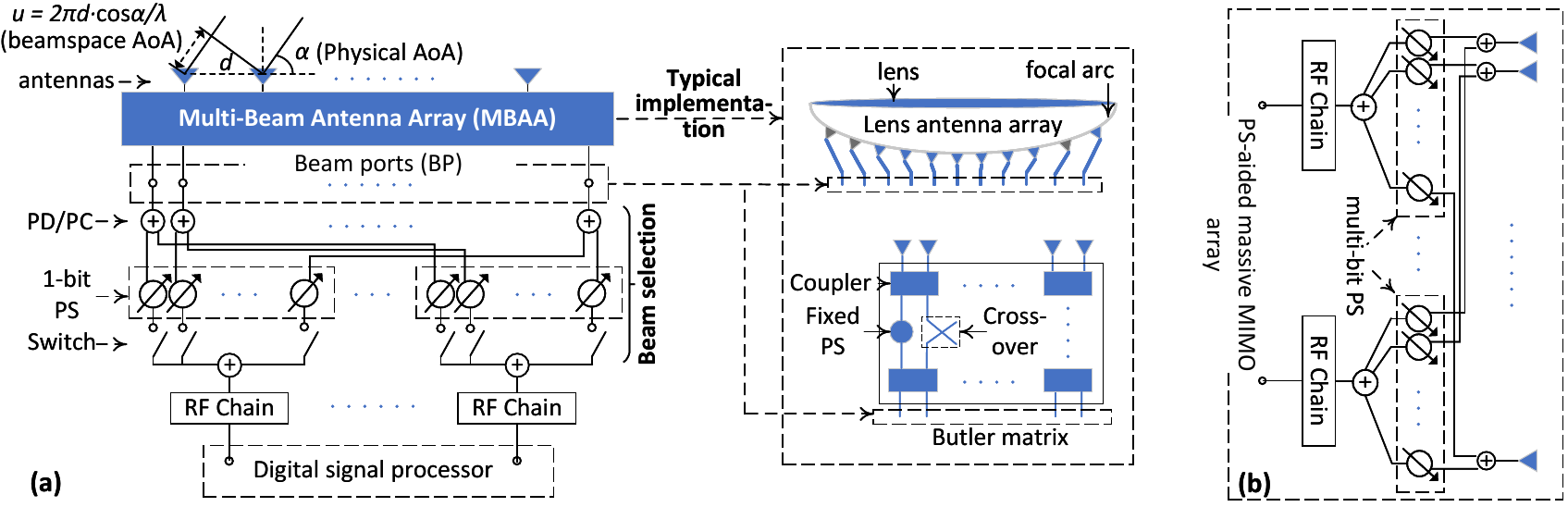}}	
	{
\caption{The schematic diagram of an MBAA-based massive MIMO transceiver is shown in (a), where PD stands for power divider, PC for power combiner, PS for phase shifters, RF for radio frequency, and DFT for discrete Fourier transform. Sub-figure (b) shows the PS-aided fully-connected hybrid array. 
}
\label{fig: antenna structure}	
}
	
\vspace{-0.5cm}

\end{figure*}

\section{Introduce Multi-Beam Antenna Arrays} \label{sec: multi-beam antenna arrays}

As depicted in Fig. \ref{fig: antenna structure}(a), an MBAA can be seen as an analog beamforming device with two interfaces. One interface consists of a number of beam ports, and the other is composed of antennas. MBAA is also a reciprocal device, meaning that it equally functions for transmitting and receiving. In the case of transmitting, the signal input into a beam port is mapped by MBAA onto radiating antennas. The specific mapping depends on the antenna design. Since the discrete Fourier transform (DFT) is most widely employed for the signal mapping \cite{Jay_Butler6G_overview}, we mainly consider the DFT-aided MBAA in this article and refer to the steered beams as the DFT beams.
	
{Let the signals on all beam ports be collected by a column vector $ \mathbf{x} $. An $ N $-size MBAA applies an $ N $-dimensional DFT on the signal and then assigns the DFT results, as denoted by $ \mathbf{y} $, to the radiating antennas. This can be expressed as $ \mathbf{y} = \mathbf{F}\mathbf{x} $, where $ \mathbf{F} $ denotes an $ N $-dimensional DFT matrix. Its $ (n,m) $-th entry can be given by $ \myDFTexp{n}{m}~(n=0,1,\cdots,N-1;~m=0,1,\cdots,N-1) $. 
Seemingly, each beam port is connected to all antenna through phase shifters with the phases configured based on a column of the DFT matrix. However, MBAAs have substantially different implementations from phased arrays.}

{As depicted in Fig. \ref{fig: antenna structure}(a), MBAA is typically implemented through lens or the circuit-type beamforming network, such as the Butler matrix \cite{Jay_Butler6G_overview}. A lens is a quasi-optical device and can be made to focus the parallel waves onto its focal arc. By placing the coupling antennas at carefully selected locations on the focal arc \cite{Jay_Butler6G_overview}, a full set of DFT beams can be steered. Differently, a Butler matrix performs the DFT beamforming by implementing the radix-$ 2 $ fast Fourier transform (FFT) using PSs and couplers. Note that PSs in the Butler matrix are fixed at pre-computed values. Regardless of the practical implementations, from the signal processing perspective, it is convenient to treat an MBAA as a ``black box''. It has equal numbers of inputs and outputs, and performs DFT on inputs to yield outputs.}

Let us briefly describe the features of the DFT beams. 
Before that, we clarify several definitions. {We refer to the angle between an incident path and the array baseline as the physical AoA, as denoted by $ \alpha $; see Fig. \ref{fig: antenna structure}(a). The propagation distance between adjacent antenna is differed by $ d\cos\alpha $, where $ d $ is the antenna spacing. This further leads to a phase difference between the two antennas, as given by $ u=2\pi d\cos\alpha/\lambda $, where $ \lambda $ denotes the wavelength. We refer to $ u $ as the beamspace AoA. 
Given $ \alpha\in[0\deg,180\deg] $ and $ d=\lambda/2 $, we have $ u\in [-\pi,\pi] $ (radian). Unless otherwise specified, the beamspace AoA is used by default in the following.}

\begin{figure}[!t]
	\centering 
	\includegraphics[width=80mm]{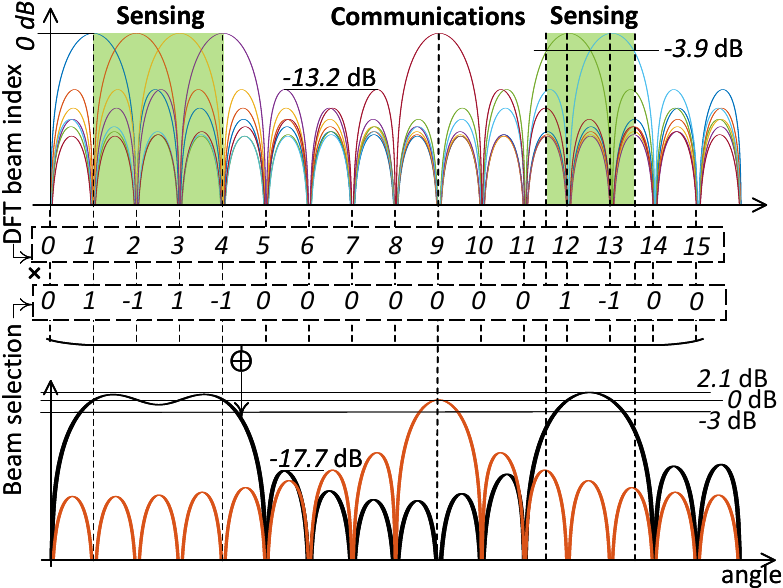}
	\caption{Beam synthesis in MBAAs. The dimension of DFT beams is $ 16 $. {Note that the beam selection shown here is only for the sensing beam with two non-contiguous wide mainlobes. The communication beam is obtained by selecting the ninth DFT beam solely.}}
	\label{fig: beam synthesis}
	\vspace{-0.5cm}
\end{figure}

{The spatial response of the $ n $-th DFT beam can be obtained by computing the discrete time Fourier transform (DTFT) of the $ n $-th column of the DFT matrix.
The magnitude of the spatial response, which is the beam pattern, has a form of $ P_n(u) = \frac{\sin(N(u-u_n))}{\sin(u-u_n)} $ \cite{KaiWu_WIPT_LAA}, where $ u_n=2\pi n/N~(n=0,1,\cdots,N-1) $. One can obtain the beam pattern expression using the geometric series formula. Features of the DFT beam are briefly summarized below. 
\begin{enumerate}
	\item At $ u_n $, $ P_n(u) $ is maximized, and hence $ u_n $ is the pointing direction of the $ n $-th DFT beam.
	
	\item Substituting $ u=u_{n'}~(\forall n'\ne n) $
	into the beam pattern expression, we obtain $ P_n(u)=0 $; hence, all other beams steer a null at the pointing direction of any DFT beam. 
	
	\item Since the nulls adjacent to the peak of the $ n $-th beam locate at $ 2\pi (n\pm 1)/N $, the first-null beamwidth is $ 4\pi/N $. 
	
	\item Near $ u=u_n $, $ (u-u_n) $ is small, and hence $  \sin(u-u_n) \approx (u-u_n) $ can be taken in the denominator of $ P_n(u) $. This turns $ P_n(u) $ into a sinc function whose first sidelobe level is well known to be about $ -13.2 $ dB. 
\end{enumerate}
One may refer to the top row of Fig. \ref{fig: beam synthesis} to 
get a more intuitive understanding of the above features.}

In mmWave/THz communications and sensing, a (super) large-scale antenna array is desired for compensating the severe propagation loss.
Consequently, we expect the number of beam ports to be large (tens to hundreds). 
To connect the beam ports with a digital processor, 
RF chains are required for frequency conversion, analog-to-digital conversion, and low-noise amplifier etc. Due to power and space limits as well as implementation difficulties, the number of RF chains that can be accommodated by a transceiver is generally much smaller than the number of beam ports.
Thus, we require some device to connect the beam ports with RF chains, leading to the so-called beam selection network in the literature.

Fig. \ref{fig: antenna structure}(a) illustrates the beam selection network. Schematically, each beam port is connected to all RF chains through a power divider/combiner (PD/PC); similarly, a power divider/combiner is also attached to each RF chain. {Each path between a beam port and an RF chain has a $1$-bit PS plus a switch, where the $ 1 $-bit PS can shift the signal phase by either $ 0 $ or $ \pi $. Therefore, if the switch is closed, the signal can go through a beam selection path with either no change or a sign reversal, leading to the beam selection weight of $ 1 $ or $ -1 $, respectively. However, if the switch is open, no signal goes through the path, making the beam selection weight become $ 0 $.}  
{In practice, only a small number of  adjacent beams have to be combined/splitted into, and the overall system can be simplified accordingly. This would simplify the implementation significantly.}
If all PSs take zeros, the beam selection network depicted in Fig. \ref{fig: antenna structure}(a) becomes solely comprised of switches, another popular structure in the literature \cite{Lens_energyEfficient_ComMag2018}.

Let us examine the power consumption of an MBAA in comparison to that of a PS-aided massive MIMO array. The latter, as mentioned earlier, has been widely employed for JCAS designs so far.  
{The diagram of the PS-aided massive MIMO array is shown in Fig. \ref{fig: antenna structure}(b).} 
We see that the main difference of the power consumed for analog beamforming in an MBAA and a PS-aided massive MIMO array 
is $ NM $ times the the power consumption difference between a 1-bit PS plus a switch and a multi-bit PS. Here, $ N $ is the number of antennas and $ M $ is that of RF chains. 

Consider the typical values of $ 30 $ milliwatt (mW) for a $ 4 $-bit PS, $ 5 $ mW for a switch and $ 5 $ mW for a $ 1 $-bit PS \cite{HybridMIMO_switchesPSs2016}. Also, assume that $ N=64 $ and $ M=8 $. Then, a PS-aided massive MIMO array can consume $ 10,240 $ mW more than an MBAA for analog beamforming, where $ 64\times 8\times (30-5-5)=10,240 $. 
The gap will be larger, if the PSs in a PS-aided massive MIMO array have more than $ 4 $ bits, or more antennas are employed.

{An advantage of
	PS-aided massive MIMO arrays is that they have more degrees of freedom in configuring the phase shifters for the beamforming that MBAA may not be capable of.} {For example, given sufficient resolutions of phase shifters, a PS-aided massive MIMO array can steer its beam to any direction in $ [0,2\pi] $; in contrast, an MBAA has a limited number of pointing directions, as illustrated earlier. 
If only one beam is selected, a loss of beamforming gain, which is up to $ -3.9 $ dB, can happen to an MBAA; see Fig. \ref{fig: beam synthesis}. As will be seen shortly, this loss can be compensated by selecting more beams, yet at a cost of widening beamwidth. }

{\section{JCAS-Critical Designs and Features of Passive MBAAs}\label{sec: relevan designs}

We proceed to review some key designs and features of MBAAs. Although they are developed recently for communications solely, they turn out to be powerful tools for multi-beam JCAS designs, as will be clear in Section \ref{sec: multi-beam JCAS framework}.} 

\subsection{Flexible Beam Synthesis} \label{subsec: beam synthesis}
Indicated by the recent overview articles on JCAS \cite{ISAC_cui2021integrating,FanLiu_overview2020TCOM,JCAS_paul2016survey}, 
one of the extensively pursued JCAS designs is the beamforming optimization for steering multiple beams, some for sensing and others for communications. While complicated optimization problems are often 
modeled and solved with non-trivial effort for the PS-aided analog beamforming, some simple beam selection can be performed in MBAAs 
for flexible and versatile beam synthesis. We use the term ``beam synthesis'' because we only select some beams of an MBAA to synthesize another beam.

Consider the design of a beam with non-zero flat mainlobes and zeros elsewhere. The number of mainlobes can be more than one and the mainlobes can be non-contiguous in $ \left[0,2\pi\right] $. Fig. \ref{fig: beam synthesis} exemplifies a desired beam with two non-contiguous mainlobes indicated by shaded rectangles. According to \cite{KaiWu_WIPT_LAA}, the beam depicted above can be approximated by:
\begin{itemize}
	\item  \textit{Identifying} the DFT beams whose superimposed mainlobes cover an angular region no smaller than the mainlobes of the desired beam; 
	
	\item {\textit{Selecting} the identified DFT beams at a single RF chain with any adjacent beams taking opposite signs.} The signs can be set using the $1$-bit PSs in the beam selection network illustrated in Fig. \ref{fig: antenna structure}(a). 
\end{itemize}
{An intuitive reason for taking opposite signs is that any adjacent DFT beams have opposite phases at the intersection of their beam patterns. Mathematically, this beam selection design can be derived based on the Fourier series expansion of the wide beam to be designed \cite{KaiWu_WIPT_LAA}.}

A graphic illustration of the beam synthesis method is given in Fig. \ref{fig: beam synthesis}. The mainlobes of DFT beams $ 1\sim 4 $ cover the angular region of the left-mainlobe in the desired beam, and DFT beams $ 12\sim13 $ cover the right-mainlobe. These beams are selected with non-zero weights, while all the other beams are assigned with zero weights, as also depicted in Fig. \ref{fig: beam synthesis}. Zero weights are set by opening the switches between the beam ports and the RF chain; see Fig. \ref{fig: antenna structure}(a). The synthesized beam, as plotted in Fig. \ref{fig: beam synthesis}(lower), has two mainlobes as desired.

We note that the beam synthesized using two DFT beams (e.g., the $ 12 $-th and $ 13 $-th beams in Fig. \ref{fig: beam synthesis}) has a mainlobe shape similar to a DFT beam, yet wider and taller. This beam turns out to be practically useful in the angle estimation, as to be seen shortly. For convenience, we call it the \textit{base-2 wide beam}. As marked in the figure, a base-2 wide beam points towards the intersection angle in the mainlobes of two adjacent DFT beams. {
	It has the beamforming gain of $ 2.1 $ dB at the intersection angle, while the original DFT beams have the gain of $ -3.9 $ dB. Thus, the base-2 wide beam improves the beamforming gain by $ 6 $ dB.}

\subsection{Angle-of-Arrival (AoA) Estimation}\label{subsec: aoa estimation}

The AoA estimation is a critical problem in both communications and sensing. 
In MBAAs, we can only perform AoA estimations using DFT beamforming results, which is known as the beamspace-domain AoA estimation. Classical estimation methods, e.g., super-resolution methods, have been extended to the beamspace domain. These methods can be adapted for MBAAs. However, they generally use DFT beams only and hence have not fully taken advantage of the benefits promised by the beam synthesis in MBAAs. For example, the synthesized base-2 wide beam can improve the beamforming gain by up to $ 6 $ dB. Next, we review a low-complexity AoA estimation method that exploits such advantage \cite{KaiWu_WIPT_LAA}. 

\begin{figure}[!t]
	\centering 
	\includegraphics[width=80mm]{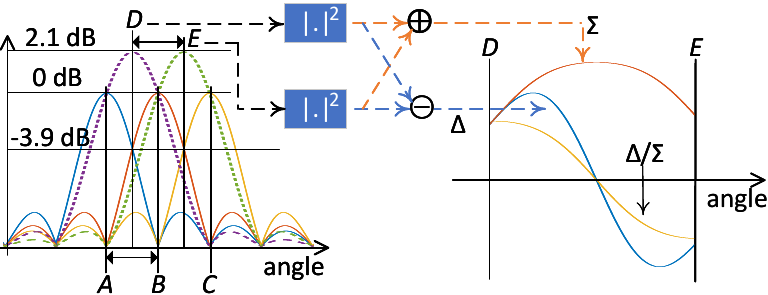}
	\caption{AoA estimation using the base-2 wide beams, where the three beams with solid curves are DFT beams, the dash curves are base-2 wide beams.}
	\label{fig: AoA curves}
	\vspace{-0.5cm}
\end{figure}

The method uses two base-2 wide beams to estimate the AoA between the pointing directions of the beams. As illustrated in Fig. \ref{fig: AoA curves}, A$ \sim $E are angles on the angle-axis. Beams-A, B and C are used to synthesize two base-2 wide beams, i.e., Beams-D and E. 
{For ease of illustration, we denote the signals received by Beam-D and Beam-E by $ x_D = g_D s $ and $ x_E = g_E s $, respectively, where $ g_D $ and $ g_E $ are the beamforming gains of the two beams towards the incident path, and $ s $ is the source signal carried by the path.}

{The method \cite{KaiWu_WIPT_LAA} first calculates the powers (absolute square) of the signals received by Beams-D and E, and then calculates the ratio of the power difference ($ \Delta $) to the power sum ($ \Sigma $). The two steps are illustrated in Fig. \ref{fig: AoA curves}. 
	Since the source signal $ s $ is common to $ x_D $ and $ x_E $, taking the ratio will suppress $ s $, leading to $ \Delta/\Sigma \approx \frac{g_{D}^2-g_E^2}{g_D^2+g_E^2} $. The approximation is due to the inevitable additive noises in $ x_D $ and $ x_E $. The ratio on the right-hand side is 
	a monotonic function of the angle between the pointing directions of Beams-D and E; see the curve tagged ``$ \Delta/\Sigma $'' in Fig. \ref{fig: AoA curves}. Moreover, the function can be analytically modeled \cite{KaiWu_WIPT_LAA}, as denoted by $ F(u) $. Then, the AoA $ u $, as the only unknown, can be uniquely solved from the equation $ F(u)\approx \Delta/\Sigma$. The less the noise power, the smaller the approximation error is, and the more accurate AoA estimation can be obtained.}

\subsection{Beam Squint Effect -- a Curse or a Blessing}
\label{subsec: beam squint}

{In theory, both communications and sensing need wide frequency bands to enable high performances, such as large communication capacity and fine sensing resolution.} In practice, wideband signals are always accompanied by some challenging problems. Beam squint is one of them \cite{FeifeiGao_wang2019overview_sp_array_massiveMIMO}. It refers to the phenomenon that the pointing direction of an antenna array changes with the working frequency. 
Consequently, the beam squint effect prevents a wideband array from 
receiving or transmitting a signal in the full frequency band towards an intended direction. 
Nevertheless, proper designs exploiting the beam synthesis reviewed earlier allow multi-beam passive antenna arrays to well handle the beam squint and even take advantage of it, as demonstrated below.

\begin{figure}[!t]
	\centering 
	\includegraphics[width=80mm]{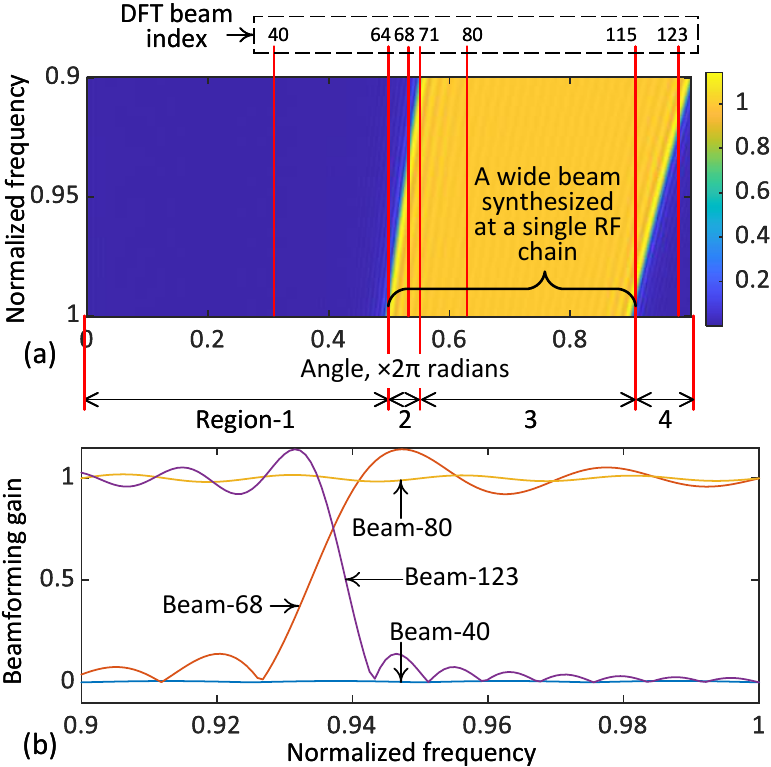}
	\caption{Illustrate the impact of beam squint in (a) and the unique spatial-frequency pattern of a wideband MBAA in (b).
	}
	\label{fig: beam squint}
	\vspace{-0.5cm}
\end{figure}

Consider a $ 128 $-dimensional MBAA that works in the frequency band $ [0.9,1] $, as normalized by the highest end. 
Apply the 
beam synthesis method reviewed in Section \ref{subsec: beam synthesis} by selecting Beams-$ 64 $ to $ 115 $ at a single RF chain. Beam-$ n $ denotes the $ n $-th DFT beam. 
The beamforming gain (in amplitude) of the synthesized beam is displayed in Fig. \ref{fig: beam squint}(a) by scaled colors. 
We see that the mainlobe of the synthesized beam
indeed covers the angular region of the selected DFT beams at the highest frequency. However, the beam squint effect gradually {shifts} the mainlobe rightwards, as the frequency decreases. 

\begin{figure*}[!t]
	\centering
	\parbox{135mm}{\includegraphics[width=13cm]{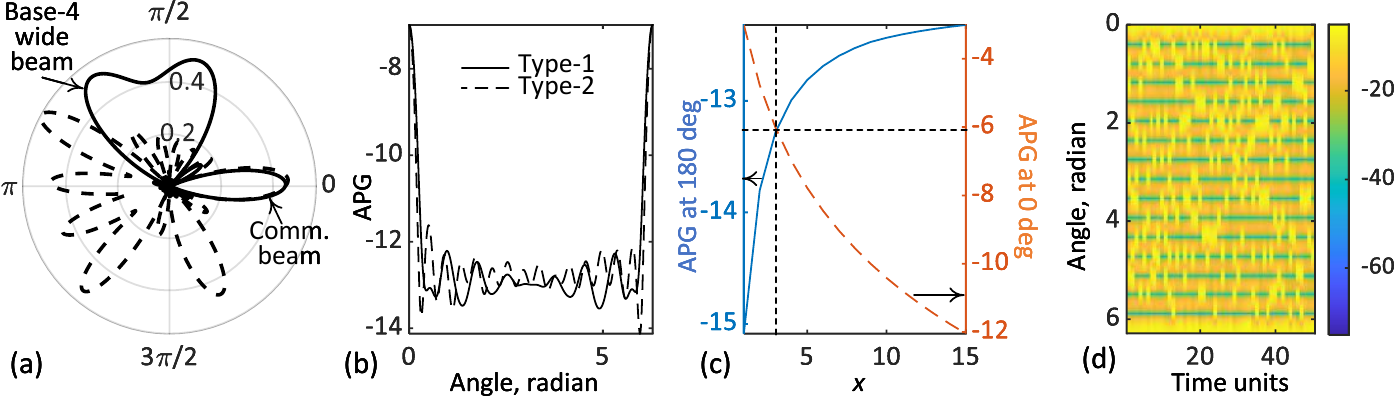}} \hfill
	\parbox{45mm}{
		\caption{Sub-figure (SF) (a) illustrates the two types of JCAS beams; their average power gains (APGs) over the whole angular region $ [0,2\pi] $ (radians) are compared in SF (b); 
			SF (c) shows the APGs of the type-1 JCAS beam towards communication (0 deg) and a sensing direction (180 deg) versus the number of simultaneously selected beams (denoted by $ x $); and SF (d) illustrates the amplitudes of the type-2 JCAS beam gains over $ [0,2\pi] $ (radians) and $ 50 $ time units. 
		}\label{fig: jcas beams}
	}

	\vspace{-0.8cm}
\end{figure*}

The maximum number of beams shifted over the whole frequency band can be quantified \cite{Kai_WidebandEffect2019JSTSP}: \textit{Beam-$ n $ at the unit frequency becomes Beam-$ x $ at the lowest normalized frequency denoted by $ \rho(< 1) $, where $ x $ is the rounding of $ n/\rho $.} For example, Beam-$ 64 $ selected at the unit frequency becomes Beam-$ 71 $ at the normalized frequency of $ 0.9 $, where $ 71 $ is the rounding of $ 64/0.9 $.
In light of the above rule, to circumvent beam squint when transmitting/receiving a full-band signal to/from the pointing direction of Beam-$ x $, we can have an RF chain select all the beams from Beam-$ n $ to Beam-$ x $, instead of Beam-$ x $ itself.

{\textit{In MBAAs, the beam squint effect can also be exploited for fast locating the AoA region of an incident path.} 
We elaborate this using the example in Fig. \ref{fig: beam squint}(a). 
Based on the beamforming gain values, we can divide $ [0,2\pi] $ into four regions, as marked in Fig. \ref{fig: beam squint}(a).
It is obvious that the beamforming gains for all angles in Region-1 (or 3) are small (or large) over all frequencies; see representative curves ``Beam-40'' and ``Beam-80'' in Fig. \ref{fig: beam squint}(b). 
Indicated by the features, the AoA of an incident path is most likely in Region-1 (or 3), if the signal power detected in the synthesized beam is all high (or all low) in the whole frequency band; otherwise, the signal may impinge from angles in Region-2 or 4.}

{The unique spatial-frequency patterns in Region-2 differentiate the region from all others. From the curve tagged ``Beam-68'' in Fig. \ref{fig: beam squint}(b), we see that the beamforming gain is small (large) in the lower (higher) end of the frequency band. Other beams in Region-2 have similar patterns, but the frequency point where the beamforming gain substantially changes is different for each beam in Region-2 \cite{Kai_WidebandEffect2019JSTSP}.
Namely, each beam in Region-2 is associated with a unique pattern depicting how beamforming gain changes over frequencies. 
If the signal power detected in the synthesized beam matches a known pattern, e.g., the curve ``Beam-68'' in Fig. \ref{fig: beam squint}(b), the incident path is most likely covered by the mainlobe of the beam.}
Similar to Region-2, there are also unique spatial-frequency patterns in Region-4, as exemplified by the curve ``Beam-123'' in Fig. \ref{fig: beam squint}(b). 
One may refer to \cite{Kai_WidebandEffect2019JSTSP} for a complete wideband beam training scheme employing the above insights.

\section{A Framework of Multi-Beam Design in JCAS}
\label{sec: multi-beam JCAS framework}

{Exploiting the designs/features illustrated above, 
we can advance the multi-beam JCAS framework recently developed in \cite{Andrew_Multibeam2019TVT} using 
MBAAs.} In the framework, a single RF chain drives two narrow beams, one for point-to-point data communications and the other performing sequential scanning for sensing. MBAAs can readily steer two such beams. They can also perform more flexible and versatile beamforming for JCAS. 
{From the transmitter perspective, two general types of JCAS beams can be steered by MBAA:}

\begin{itemize}
	\item \textit{Type-1. Regular beam synthesis for sensing:} While a single DFT beam is selected for communications, another beam, either narrow or wide, can be synthesized for probing a sensing volume. An example of the type-1 JCAS beam is demonstrated by the solid curve in Fig. \ref{fig: jcas beams}(a), where the communication beam points towards the zero degree, and a base-4 wide beam is synthesized for sensing using the method in Section \ref{subsec: beam synthesis}.

	\item \textit{Type-2. Random beam selection for sensing:} The sensing beams are randomly selected from the DFT beams that are not assigned for communications.
	An example is given in Fig. \ref{fig: jcas beams}(a). With the zero-th DFT beam kept for communications, a random selection of four out of other beams is performed, leading to the beam pattern depicted by the dash curve.  
	 
\end{itemize}

{Utilizing the beam synthesis method illustrated in Section \ref{subsec: beam synthesis}, the type-1 JCAS beam can be exploited 
	for a bi-/multi-section spatial search for target localization; see \cite{KaiWu_WIPT_LAA}, which has a logarithmic time complexity. In contrast, the conventional sequential beam scanning has a linear time complexity \cite{Andrew_Multibeam2019TVT}.
Moreover, enabled by the AoA estimation method reviewed in Section \ref{subsec: aoa estimation}, the 
bi-/multi-section spatial search can stop at the base-2 beams, and the received signals can be used for the accurate AoA estimation.
In addition, when wideband signals are available, the beam squint effect, as described in Section \ref{subsec: beam squint}, can be exploited for ensuring the type-1 JCAS beam to cover the sensing direction with the whole frequency band.}

For the type-2 JCAS beam, the random spatial sampling is performed, and hence the compressive sensing techniques may be applied for JCAS. 
Note that specific JCAS schemes using the two types of beams have not been developed in the literature yet. 
More details of the research challenges/opportunities regarding this will be provided in Section \ref{sec: challenges of multi-beam antenna arrays}. Here, we unveil more features of the two JCAS beams.

\textit{With the same number of DFT beams selected, the two types of JCAS beams have similar average power gain over the sensing region regardless of their substantially different sensing beams.} The average power gain is the squared beamforming gain averaged over a period of observation time.
To illustrate the point, we fix the communication direction at the zero degree and perform sensing in the whole angular region $ [0,2\pi] $. 
A total of $ 16 $ DFT beams are adopted, and four beams are selected for sensing in both types of JCAS beams. The indexes of the four beams in the type-1 JCAS beam cyclically increase by one for each new time unit; for example, [12,13,14,15] for the first time unit, [13,14,15,0] for the second and so on. For the type-2 JCAS beam, the indexes of the four DFT beams selected for sensing are randomly taken from $ [0,15] $ in each time unit. Calculating the average power gains of the two types of JCAS beams over $ 1,000 $ time units, we obtain the two curves in Fig. \ref{fig: jcas beams}(b). We can see the similar average power gain distribution of the two types of beams.

\textit{For both types of JCAS beams, there is a non-trivial trade-off between power gains for communications and sensing caused by the number of simultaneously selected beams,} as denoted by $ x $ for convenience. An intuitive reason is that all the selected beams connected to the same RF chain share the signal power evenly through a power divider; see Fig. \ref{fig: antenna structure}(a). To illustrate the trade off, we 
re-perform the simulation for Fig. \ref{fig: jcas beams}(b) with $ x $
taking from $ 2$ to $15 $. Fig. \ref{fig: jcas beams}(c) plots the power gains in the directions of the communication user (zero degrees) and of sensing ($ 180 $ degrees), as $ x $ increases. We see that the power gain for sensing (communications) increases (decreases) with $ x $. Interestingly, the two curves have substantially different changing rates against $ x $, particularly around their intersection. Thus, there may be an optimal trade-off to be discovered.

\textit{While the type-1 JCAS beam exposes the communication information to potential eavesdroppers, the type-2 JCAS beam naturally enhances the physical-layer secrecy.}
{In the single-RF chain case considered here, the same signal is used for communications and sensing. Thus, the type-1 JCAS beam can expose communication signals to potential eavesdroppers, due to its regular (hence traceable) beam scanning.} 
In contrast, the random selection in the type-2 will scramble the amplitudes and phases in non-communication directions, which makes it challenging, if not impossible, for an eavesdropper to recover communication signals. Fig. \ref{fig: jcas beams}(e) shows the absolute gain of the type-2 JCAS beam over the whole angular region and $ 50 $ time units (each with a random selection of four DFT beams for sensing and the zero-th DFT beam fixed for communications). It is clear that except in the communication direction (the zero degree), the absolute gain randomly changes over other angles and time units; namely, the communication information is fully scrambled in non-communication directions. The phases of the beamforming gains are also random, which is not shown here due to the limit on the number of figures.

\section{Challenges and Opportunities
 }\label{sec: challenges of multi-beam antenna arrays}

{While the flexible and versatile beamforming of MBAAs demonstrated above using the existing techniques reviewed in Section \ref{sec: relevan designs}, other key aspects of JCAS, including waveform design, receiving algorithms, JCAS beam selections etc., have not been explored yet.} Challenges and potential solutions in these topics are highlighted  next.  

\subsubsection{Waveform Design}
As reviewed in Section \ref{sec: relevan designs}, MBAAs enjoy great flexibility in beam synthesis. Thus, we may use the methods reviewed earlier to synthesize analog beams approximating desired beam patterns and then focus on refining the final beam in the digital domain only. 
In addition, the unique spatial-frequency relation in MBAAs, as illustrated in Section \ref{subsec: beam squint}, can be exploited to perform waveform design in the joint spatial-/time-/frequency-domain. 

{When performing JCAS waveform design, we shall keep in mind the non-trivial trade-off between communications and sensing caused by the power allocation, as illustrated in Section \ref{sec: multi-beam JCAS framework}. Moreover, the mutual interference between the two functions always has to be kept low. A joint space-time-frequency precoding may be exploited to generate two orthogonal spaces for the waveform design of the two functions. In addition, the communication secrecy constitutes another trade-off between communications and sensing in MBAAs. As observed  
in Section \ref{sec: multi-beam JCAS framework}, more randomness in sensing waveform can help protect communication information from eavesdropping. This, however, can make the sensing receiver design more complicated, as to be discussed shortly. It is noteworthy that an increasing number of JCAS techniques exploit the sensing results for enhancing the communication secrecy\footnote{{Interested readers are referred to: \url{https://www.comsoc.org/publications/best-readings/integrated-sensing-and-communication-isac}, for a summary on the related literature.}}. This may be a way to relieve the secrecy constraints during the waveform design.}

MBAAs also bring new improvements/opportunities to the radar-centric (RC) JCAS that embeds communication information into existing radar waveforms \cite{DFRC_JRC_AutonomousVehicle_2020_SPmag}. 
\textit{First}, the multi-beam nature allows multiple communication users to be served with different beams, which may be exploited for reducing inter-user interference. 
\textit{Second}, MBAAs offer a new way of information modulation, i.e., beam hopping (BH), to further improve the data rate. An example of BH is given below. 

{Consider that there are three dominant paths covered by different beams in a point-to-point communications. We can select some of the beams for communications and use the variety in the beam selection to convey more information bits. In this example, we have $ 7(=C_3^1+C_3^2+C_3^3)  $ combinations of beams, where $ C_x^y $ denotes the the number of combinations of taking $ y  $ out of $ x $ items at a time. 
	Out of the seven beam combinations, four best ones, which can be determined based on some metric, e.g., maximum signal-to-interference ratio, can be used for communications. 
The varying information bits make the beams hop over communication symbols, hence the name BH.}
Similar to BH, the antenna selection, another type of index modulation, has been used for information modulation in RC-JCAS \cite{DFRC_JRC_AutonomousVehicle_2020_SPmag}. These designs \cite{SpatialModulaiton_JianSong2021_comMag,DFRC_JRC_AutonomousVehicle_2020_SPmag} may enlighten the development of BH-based RC-JCAS in MBAAs.

 \begin{figure*}[!t]\footnotesize
	\captionof{table}{A Summary of Challenges and Opportunities Arisen from Applying Passive MBAAs to JCAS. References (Ref.) from other applications/background, yet implying potential solutions, are also listed.}
	\label{tab: challenges and solutions}
	\centering
	\includegraphics[width=16cm]{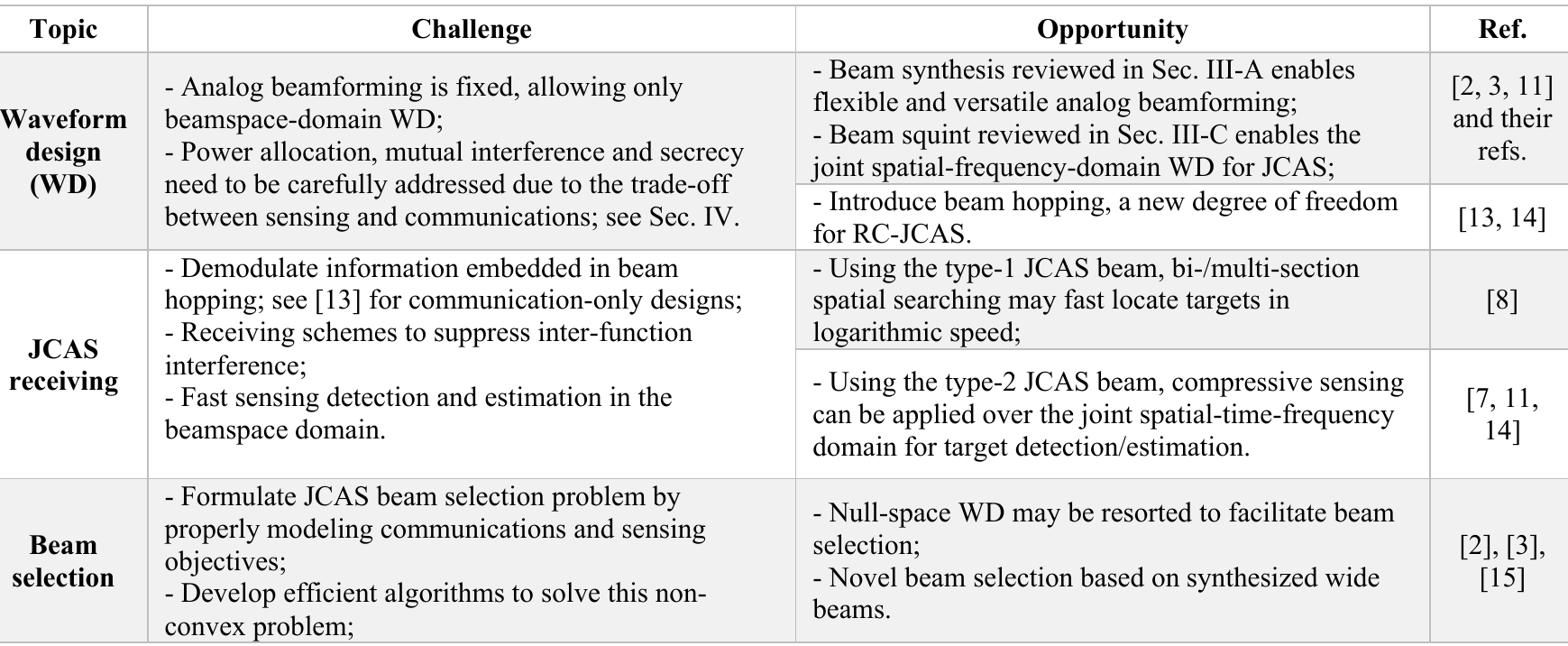}
	\vspace{-0.5cm}
\end{figure*}

\subsubsection{JCAS Receiving Design}
Corresponding to the new potential information modulation through BH, new algorithms have to be developed for demodulating the BH-carried communication information. The methods and analysis delivered in \cite{SpatialModulaiton_JianSong2021_comMag} may serve as a good reference, though the work deals with the beam index (de-)modulation for data communications.

For sensing, we can steer regular beams like in the type-1 JCAS beam. A bi-section or multi-section fast spatial searching can be performed by coordinating the sensing beams at the transmitter and the receiver. An example has been given in Section \ref{sec: multi-beam JCAS framework}. One may refer to \cite{KaiWu_WIPT_LAA} for a complete multi-section spatial searching scheme in the narrowband case. One may also refer to \cite{Kai_WidebandEffect2019JSTSP} for the wideband case, where the beam squint effect illustrated in Section \ref{subsec: beam squint} is applied.

For the sensing receiving, we can also employ the random sensing beam in the type-2 JCAS beam. As mentioned in Section \ref{sec: multi-beam JCAS framework}, the compressive sensing techniques can be employed for spatial localization. 
A similar scheme is developed for the channel estimation in MBAA-aided massive MIMO communications \cite{Lens_energyEfficient_ComMag2018}. 
For sensing, we require not only spatial information but also the target ranges, velocities and polarities etc. Thus, more investigations are required either to extract the time-frequency-domain target information based on the spatial-domain compressive sensing results or to reformulate multi-dimensional compressive sensing problems to factor in all key target parameters; see \cite{JCAS_paul2016survey,SpatialModulaiton_JianSong2021_comMag} and their references.

 \subsubsection{JCAS Beam Selection}
 Beam selection has been a hot topic in MBAA-based data communications, where the design goal is generally to maximize the communication sum rate. {The beam selection network, as illustrated in Fig. \ref{fig: antenna structure}(a), only has limited discrete states, making the beam selection problem analytically intractable. The non-convex objective function and feasible region make the beam selection problem NP-hard. 
Existing solutions, e.g., \cite{Lens_beamSelec_TCOM2015_Masouros,Lens_energyEfficient_ComMag2018}, mainly resort to the sub-optimal incremental or decremental search, having a computational complexity proportional to the third power of the overall beam number \cite{Lens_beamSelec_TCOM2015_Masouros}. 
More research efforts are expected to further reduce the beam selection complexity.}

A key reason making the beam selection problem difficult is that the user channels are not completely orthogonal in the beamspace domain. 
Thus, a major task of beam selection is to reduce inter-user interference without losing too much signal power. Now, with radar sensing added, we have to further deal with inter-function interference. {If the same waveform is used for communications and sensing, we can extend the previous beam selection algorithms \cite{Lens_beamSelec_TCOM2015_Masouros} to JCAS by treating sensing beams as communication beams. A lower bound on the average power gain for sensing may be posed as a constraint to ensure satisfactory sensing performance. 
If different waveforms are allowed, we may consider designing sensing waveforms in the null-space of the communication waveforms \cite{DFRC_JRC_AutonomousVehicle_2020_SPmag}, or vice versa, intertwining waveform design and beam selection problems. }

\section{Conclusions}\label{sec: conclusions}

{We intuitively illustrated what an MBAA is and also several JCAS-critical designs and features of the array. 
Exploiting those designs/features, we demonstrated how MBAAs can advance a recently developed multi-beam JCAS framework using their flexible and versatile beamforming capabilities. 
Furthermore, we elaborated the challenges of several key aspects for comprehensively applying MBAAs to JCAS. We suggested potential solutions to the challenges by highlighting existing handy designs in different applications, as summarized in Table \ref{tab: challenges and solutions}.
The multi-beam array is promising for enabling easy-to-design and energy-efficient green JCAS systems.}

\bibliographystyle{IEEEtran}
\bibliography{IEEEabrv,./bib_JCAS.bib}

\vspace{0.1cm}

{\footnotesize
	\noindent \textbf{Kai Wu} (Kai.Wu@uts.edu.au, \textit{Member}, IEEE) is a research fellow with the University of Technology Sydney (UTS), Sydney, Australia. He received PhD from UTS in 2020.
	
	\vspace{0.1cm}
	
	\noindent \textbf{Dr J. Andrew Zhang} (Andrew.Zhang@uts.edu.au, \textit{Senior Member}, IEEE) is an associate professor with UTS. He received his PhD from the Australian National University in 2004.
	
	\vspace{0.1cm}
	
	\noindent
	\textbf{Xiaojing Huang} 
	(Xiaojing.Huang@uts.edu.au, \textit{Senior Member}, IEEE) is a professor with UTS. He received his PhD from Shanghai Jiaotong University in 1989.
	
\vspace{0.1cm}
	
	\noindent
	\textbf{Robert Heath} (rwheathjr@ncsu.edu, Fellow, IEEE) is a professor with North Carolina State University, USA. He received his Ph.D. from Stanford University, Stanford, CA, in 2002.
	
	\vspace{0.1cm}
	
	\noindent
	\textbf{Y. Jay Guo} 
	(Jay.Guo@uts.edu.au, \textit{Fellow}, IEEE) is a distinguished professor with UTS. He received his PhD from Xi’an Jiaotong University in 1987.	
}

\end{document}